\documentclass[11pt,twoside]{article}
\usepackage{asp2004}
\usepackage{psfig}
\usepackage{epsf}
\usepackage{graphics}
\usepackage{lscape}
\makeindex 
\begin{document}
\title{Interested in observing TrES-Her0-07621?}
\author{O.~L.~Creevey$^{1,2}$, 
Timothy M.~Brown$^{1}$, 
Sebastian Jim\'enez-Reyes$^{2}$\\
and 
Juan Antonio Belmonte$^{2}$}
\affil{$^1$High Altitude Observatory, NCAR, Boulder 80307, CO USA}
\affil{$^2$Instituto de Astrof\'isica de Canarias, E-38200, Tenerife, Spain}

\begin{abstract}                                                              

TrES-Her0-07621 is a recently discovered detached M Dwarf 
eclipsing binary system.  We present some follow-up observations of this
system including new minima times and a refined orbital period.
We have also obtained better estimates of the stellar radii and inclination.

\end{abstract}

\section{Introduction\label{sec:intro}} 

The detached M Dwarf eclipsing binary TrES-Her0-07621 was published recently
as an addition to the few such binary systems previously known
\citep{creevey05} (see references within, C05 henceforth).
This object is interesting because M Dwarfs are abundant in our galaxy, 
and yet are poorly understood.
The binary nature allows us to determine accurate stellar radii and
masses.
Previously we obtained  precise mass estimates of 
both components of TrES-Her0-07621 
(Hobby-Eberly Telescope, McDonald Observatory).
However, due to the 11\arcsec/pixel resolution of the CCD detectors, our
photometric time series 
is contaminated by 
a nearby non-variable star, thus providing limitations on the photometric
accuracy and precision.  We present some new  
photometric results based on recently observed non-contaminated time series. 
\index{TrES-Her0-07621} \index{M Dwarf}

\section{Observations and Results\label{sec:observations}}
We obtained observations with the {\it IAC80} and {\it TCS}
at the Observatorio del Teide, Tenerife
and with {\it NOT} at the Observatorio Roque de los Muchachos, La Palma.
The data were reduced using standard reduction procedures and we performed
aperture photometry to obtain the time series.

We observed a total of 10 light curves in various filters at 6 different 
minimum epochs.  We tested to make sure that 
there were no phase shifts of minimum with filter (we also found no 
depth dependance).
Each of these light curves were fit to a binary system model to 
obtain the eclipse
minima.  To obtain a more accurate photometric period, we used the first 
observed eclipse by STARE (2003) and the quoted
HJD from C05, along with the new minima.  
We fit a straight line to the HJD of observed minimum times as a function
of elapsed orbital cycle.
The slope of this line gave a period of
1.120804 ($\pm$ 0.000012).
Table 1 shows the new eclipse minima for both primary and secondary eclipses.

\begin{table}[t]
\center{\caption{Eclipse Minima}}
\label{table:observations}
\footnotesize
\begin{center}
\begin{tabular}{lcccc}
\noalign{\smallskip}
\tableline
\noalign{\smallskip}
HJD & Error & Cycle No. & Eclipse & Observed with\\
\tableline
\noalign{\smallskip}    
2452766.5161  & 0.0001  & 0 & 2& $R$ STARE\\
2453139.74951 & 0.000001  & 333 & 2& $R$ IAC80\\
2453503.4368 & 0.0008 & 657.5  & 1& $R$ NOT\\
2453503.4362 & 0.001 & 657.5 & 1& $V$ NOT\\  
2453523.6108 & 0.001 & 675.5  & 1& $J$ TCS\\
2453527.5333 & 0.001 & 679  & 2& $J$ TCS\\
2453536.4999 & 0.0004 & 687  & 2& $R$ NOT\\
2453536.4995 & 0.0004 & 687 & 2& $V$ NOT\\
2453554.4313 & 0.0006 & 703  & 2& $R$ IAC80\\
2453554.4306 & 0.0006 &703  & 2& $I$ IAC80\\
2453555.5533 & 0.0004 & 704 & 2& $R$ IAC80\\
2453555.5529 & 0.001 & 704 & 2& $I$ IAC80\\ 
\noalign{\smallskip}
\tableline
\end{tabular}
\end{center}
\end{table}

We used the data from NOT to model the binary system (Figure 1).
With this higher precision photometry we have obtained better estimates for
both radii, inclination and effective temperature ratio (Table~2.)

\begin{figure}[ht!]

\centerline{
\psfig{figure=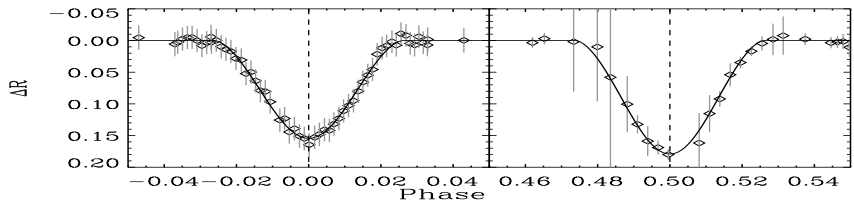,angle=0,height=2.8cm,width=12.cm}}
\caption[]{
Model fit to the data.  We show the light curve fit of both eclipses,
the data were observed with NOT using Bessell $R$ filter.}
\label{fig:fig1creevey2} 
\end{figure}

\begin{table}
\center{\caption{Updated Component and System Parameters}}
\label{table:param}
\begin{center}
\begin{tabular}{lrclr}
\noalign{\smallskip}
\tableline
\noalign{\smallskip}
\noalign{\smallskip}
\tableline
\noalign{\smallskip}
R$_A$(R$_{\odot}$) & 0.449(0.030) && R$_B$ &  0.449(0.030) \\
M$_A$(M$_{\odot}$) & 0.493(0.003) && M$_B$ & 0.489(0.003)\\
$i$($^{\circ}$)& 83.10(0.30)  && T$_B$/T$_A$  & 0.97(0.02)  \\
\noalign{\smallskip}
\tableline
\end{tabular}
\end{center}
\end{table}

\section{Conclusions\label{sec:concl}} 
We have provided new minima times and a refined photometric period for the
eclipsing binary system TrES-Her0-07621.  
We have also obtained better radii estimates. 
We hope that this system will contribute to learning more about low mass
stars, there is a lack of such systems observed and so  any new ones are
not only 
of benefit 
but necessary in order to advance in this area of stellar physics.

{}

{\small Acknowledgements\\
Thank you to IAC80 and TCS at the Observatorio del Teide, Tenerife
and NOT at the Observatorio Roque de los Muchachos, La Palma (Spain)
for observing time and their knowledgeable staff: 
John Telting,
Erik Stempels (NOT),
Jos\'e Miguel Gonz\'alez P\'erez, 
Alfred Rosenberg,
Luis L\'opez Mart\'in (OT).
Thanks to Peter Hammersley for help with the IR data processing.
Thanks to Travis Metcalfe and Chris Sterken for 
their suggestions regarding this work.}

\end{document}